\documentclass[conference]{IEEEtran}
\IEEEoverridecommandlockouts
\usepackage{cite}
\usepackage{amsmath,amssymb,amsfonts}
\usepackage{bm}
\usepackage{algorithmic}
\usepackage{graphicx}
\usepackage{textcomp}
\usepackage{xcolor}
\usepackage{capt-of}  % <---
\usepackage{cuted}    % <===
\usepackage{multirow}
\usepackage{lipsum}
\usepackage{hyperref}
\usepackage{caption}
\usepackage{subcaption}
\def\BibTeX{{\rm B\kern-.05em{\sc i\kern-.025em b}\kern-.08em
    T\kern-.1667em\lower.7ex\hbox{E}\kern-.125emX}}
\begin{document}

\title{A Spatially Separable Attention Mechanism for massive MIMO CSI Feedback\\}

\author{\IEEEauthorblockN{Sharan Mourya}
% \IEEEauthorblockA{\textit{IIT, Hyderabad} \\
% Hyderabad, India \\
% sharan.mourya@5g.iith.ac.in}\\
\and
\IEEEauthorblockN{SaiDhiraj Amuru}
% \IEEEauthorblockA{\textit{IIT, Hyderabad} \\
% Hyderabad, India \\
% asaidhiraj@ee.iith.ac.in}
\and
\IEEEauthorblockN{Kiran Kumar Kuchi}
% \IEEEauthorblockA{\textit{IIT, Hyderabad} \\
% Hyderabad, India \\
% kkuchi@ee.iith.ac.in}
}

\maketitle

\begin{abstract}
Channel State Information (CSI) Feedback plays a crucial role in achieving higher gains through beamforming. However, for a massive MIMO system, this feedback overhead is huge and grows linearly with the number of antennas. To reduce the feedback overhead several compressive sensing (CS) techniques were implemented in recent years but these techniques are often iterative and are computationally complex to realize in power-constrained user equipment (UE). Hence, a data-based deep learning approach took over in these recent years introducing a variety of neural networks for CSI compression. Specifically, transformer-based networks have been shown to achieve state-of-the-art performance. However, the multi-head attention operation, which is at the core of transformers, is computationally complex making transformers difficult to implement on a UE. In this work, we present a lightweight transformer named STNet which uses a spatially separable attention mechanism that is significantly less complex than the traditional full-attention. Equipped with this, STNet outperformed state-of-the-art models in some scenarios with approximately $1/10^{th}$ of the resources.
\end{abstract}

\begin{IEEEkeywords}
STNet, CSI Feedback, Transformers, Self-Attention, Massive MIMO, TransNet, CSIFormer, CLNet.
\end{IEEEkeywords}

\section{Introduction}
A massive MIMO system is equipped with hundreds of antennas that facilitate increased throughput with reduced BLER (Block Error Rate). Real-time channel state information (CSI) at the base station (gNB) plays an important role to meet the promises offered by massive MIMO. CSI at gNB allows the base station to perform beamforming and serve multiple users at once with minimal interference. The difficulty in this is that the channel experienced by the user equipment (UE) in the downlink has to be measured by UE and send it back to gNB in real-time which is not very convenient given the limited amount of resources at the UE end and the huge amount of overhead caused by CSI on the uplink. In order to deal with this problem, CSI compression methods were introduced where we compress the channel matrix at UE to reduce the feedback overhead and power consumption. \par
A Compressive Sensing (CS) based CSI feedback in FDD systems was studied in \cite{CS} where it was achieved by using 2-D Discrete Cosine Transform (DCT) or Karhunen-Loeve Transform (KCT). By exploiting sparsity, CS facilitates efficient data sampling at much lower rates than determined by the Nyquist theorem. However, this method assumes the channel matrices to be sparse which is not always the case. For efficient compression, the spatial correlation characteristics of the channel matrix have to be exploited which was proposed in \cite{PCA} by using a principal component analysis (PCA). Even in this method, sparsity of the channel matrix in some basis is assumed for efficient compression but channels don't always have an interpretable structure. \par
In order to overcome this, a data-driven approach is chosen over an algorithmic-driven one and the usage of deep learning in CSI compression has taken over in recent years. CSINet \cite{csinet} introduced a Convolutional Neural Network (CNN) based Variational Auto-Encoder (VAE) to the compression problem. This tremendously outperformed  all the traditional CS-based methods. Inspired by this, another model was developed \cite{csinet+} with a larger receptor size, i.e., kernel size, of the CNN to better capture the spatial correlation in the angular-delay domain. However, the variability of the channel sparsity with the scenario means that a fixed receptor size is not sufficient to capture the correlation. Hence, a multiple-resolution CNN with varying receptor sizes was introduced by CRNet \cite{crnet}. In order to focus the resources more on highly correlated areas and less on less correlated areas, it is useful to employ an attention mechanism on the CNNs which was first introduced by Attention-CSINet \cite{attnnet} that performed better in outdoor scenarios where the variability of correlation is more dominant. In all these methods, real and imaginary parts of the channel matrices are treated separately which is not efficient in capturing the correlation in the angular domain as the complex number as a whole contains the phase information. To overcome this, a simple approach to combining real and imaginary values of a channel matrix was introduced by CLNet \cite{clnet} which outperformed several models. CLNet is also computationally less complex compared to other methods.\par
\begin{figure*}
\includegraphics[width=\linewidth]{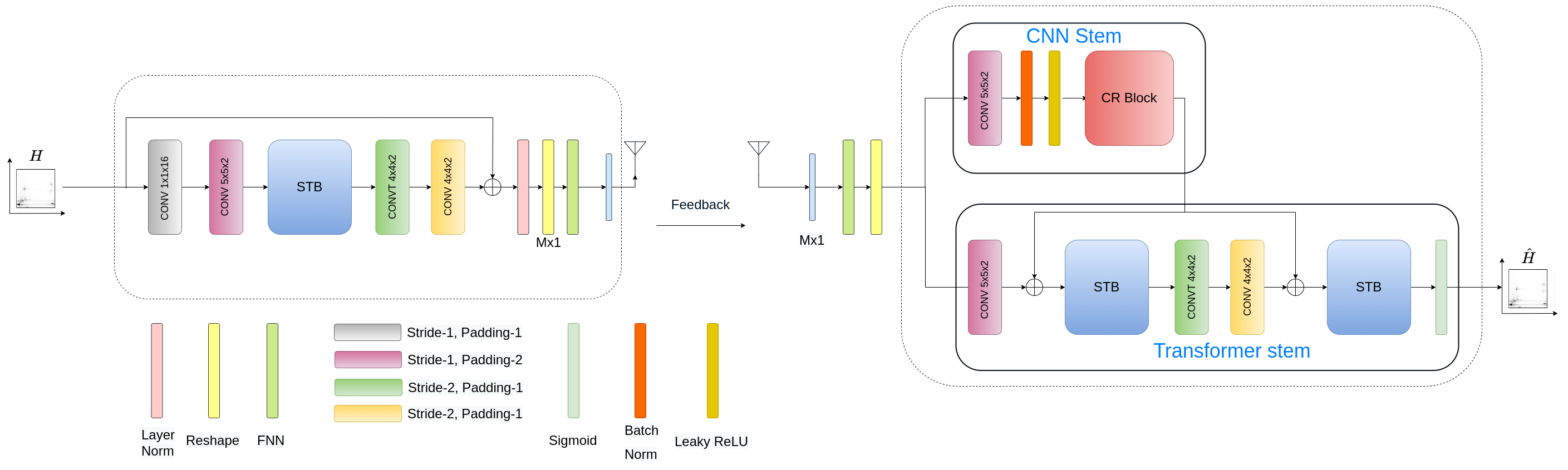}
\captionof{figure}{Proposed encoder-decoder architecture for CSI feedback aka STNet. "CONV" represents a convolutional layer and "CONVT" represents a transposed convolutional layer. The encoder consists of a few CONV and CONVT layers with a spatially separable attention transformer block "STB". The decoder consists of two stems: the CNN stem and the transformer stem. CR Block is a multi-resolution CNN block proposed in CRNet as shown in Fig.{~\ref{crblock}} }
\label{net}
\end{figure*}
So far, CNNs are used for feature extraction in all the models. A transformer \cite{transformer}-based architecture with a full attention mechanism was first studied in \cite{empower} that was not very competitive compared to the state-of-the-art models. A two-layer transformer architecture named TransNet \cite{transnet} was introduced that outperformed several models by a significant amount but the computational complexity of TransNet was very high and not practically affordable. Another transformer-based model with locally grouped (windowed) self-attention was studied by CSIFormer in \cite{csifromer}. Although this has low complexity compared to TransNet, the performance was sacrificed. \par

In this work, we introduce a spatially separable attention mechanism \cite{ssa} that can achieve state-of-the-art performance with very less computational complexity. We also introduce a hybrid two-stem approach in the decoder that combines CNN with a transformer for better channel reconstruction \cite{csformer}. We then validate the performance of our model on the COST2100 dataset \cite{cost}.

\section{System Model}

In this work, we consider a Frequency Division Duplex (FDD) system with $N_{t}$ antennas at the base station (gNB) and 1 antenna at the user equipment (UE) such that $N_{t}\gg 1$. This employs Orthogonal Frequency Division Multiplexing (OFDM) with $\Tilde{N}_{c}$ sub-carriers. The received signal at UE on the $n^{th}$ sub-carrier can be expressed as
\begin{equation}
    y_{n} = \textbf{$\bm{\Tilde{h}_{n}}^{H}$}\bm{\Tilde{v}_{n}}x_{n} + \textbf{$w_{n}$},
\end{equation}
where, $\bm{\Tilde{h}_{n}} \in \mathbb{C}^{N_{t}\times 1}$, $\bm{\Tilde{v}_{n}} \in \mathbb{C}^{N_{t}\times 1}$, $x_{n} \in \mathbb{C}$ and $w_{n} \in \mathbb{C}$ represent channel vector, precoding vector, symbol transmitted and additive noise on the $n^{th}$ sub-carrier.
The overall channel matrix is of the dimension $\Tilde{N}_{c} \times N_{t}$ and is expressed as
\begin{equation}
    \bm{\Tilde{H}} = \big[\bm{\Tilde{h}_{1},\Tilde{h}_{2},\cdots,\Tilde{h}_{\Tilde{N_{c}}}} \big]^{H}.
\end{equation}
The total number of feedback elements is $2N_{t}\Tilde{N}_{C}$ (for both real and imaginary parts of the channel) which is huge and impractical in real scenarios considering $N_{t}=32,64,\hdots$ and $\Tilde{N}_{C} = 1024,2048,\hdots$ for a massive MIMO system. So, we reduce the overhead by making the channel matrix sparse. Specifically, we achieve this by transforming it into angular-delay domain \cite{csinet} as follows
\begin{figure}
\centering
\includegraphics[width=2in,height = 1.2in]{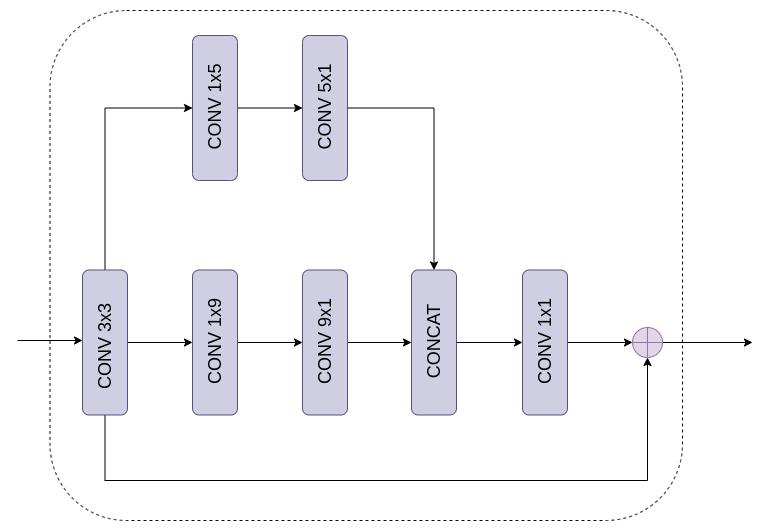}
\captionof{figure}{CR Block as proposed in CRNet architecture. It consists of two paths with different kernel sizes that are concatenated (represented by "CONCAT" block) at the end and combined using a 1x1 convolutional layer.}
\label{crblock}
\end{figure} 
\begin{equation}
    \bm{\bar{H}} = \bm{F_{d}\Tilde{H}F_{a}^{H}},
\end{equation}
where, $\bm{F_{d}}$ and $\bm{F_{a}}$ are 2-D DFT matrices of dimensions $\Tilde{N}_{c}\times \Tilde{N}_{c}$ and $N_{t} \times N_{t}$ respectively. In the delay domain, the time delay between multipath arrivals lies within a limited period. Using this, we can truncate the matrix $\bm{\bar{H}}$ by only keeping the first $N_{c}$ rows where $N_{c}$ is chosen such that remaining entries of $\bm{\bar{H}}$ are close to zero \cite{csinet}. We define this truncated matrix as $\bm{H}$ that has dimensions $N_{c}\times N_{t}$. Also, we split this matrix into real and imaginary parts and combine them as a third dimension similar to RGB channels of an image. With this, the overall feedback overhead becomes $2N_{c}N_{t}$ which is significantly smaller than earlier as $N_{c}$ will only be a fraction of $\Tilde{N}_{c}$ (total number of sub-carriers). 
\par
Now that we have the sparsified channel matrix, $\bm{H}$, it is sent into the encoder-decoder architecture as shown in Fig.{~\ref{net}} where $\bm{H}$ is compressed into a 1-D vector of dimension $M \times 1$. Here, we define compression ratio as $\gamma = \frac{M}{2N_{c}N_{t}}$. This compressed channel matrix is sent back to gNB from UE on the uplink. gNB then decodes this fedback signal as $\bm{\hat{H}}$. This encoding and decoding process is defined as follows
\[s = f_{e}(\bm{H}) \quad \& \quad \bm{\hat{H}} = f_{d}(s),\]
where $f_{e}$, $f_{d}$ denote the functions of the encoder and decoder, respectively. $s$ is the compressed code word and $\bm{\hat{H}}$ is the estimated channel matrix by the model.

\section{Architecture}
% \begin{strip} % <--- defined in "cuted"
% \includegraphics[width=\linewidth]{model.png}
% \captionof{figure}{Encode-Decoder Architecture}
% \label{Fig:image label}
% \end{strip}
In this section, we describe a high-level overview of how $f_{e}$ and $f_{d}$ of our proposed model STNet\footnote{Source code of this paper: \url{https://github.com/sharanmourya/Pytorch\_STNet}} are designed.
% \subsection{Two-stem approach}
%The encoder mainly consists of a CNN layer in the beginning layers to extract the local features as early convolutions improve the transformer's stability and performance \cite{early}. This is followed by a Spatially Separable Attention Transformer Block (STB) that captures the long-range correlation between antennas. In this, we introduce a spatially separable attention mechanism which is described in detail in the later sections. %The STB is followed by a couple of CNN layers and a linear layer i.e., a fully connected layer to project the code word onto a lower dimension suitable for transmission which is then recovered by the decoder.%
Transformers are traditionally designed to capture global context using a global self-attention mechanism which makes them highly efficient in modeling high-level semantics that may be sufficient for a classification task. For example, in our model Spatially Separable Attention Transformer Block (STB) captures the long-range correlation between antennas. But, channel reconstruction also requires low-level details in order to minimize the reconstruction error. These low-level details are better captured by CNNs which also provide better generalization due to their spatial invariance. So in order to get the best of both worlds, we use a hybrid approach in our decoder design with two stems \cite{csformer}, one consisting of a transformer and the other consisting of CNNs (Fig.{~\ref{net}}). %The transformer stem consists of CNNs and STBs. The CNN stem consists of a CNN layer followed by a CR Block \cite{crnet} (Fig.{~\ref{crblock}}) whose output is then added to the inputs of the STBs in the transformer stem making it easier for the transformer to extract antenna correlation.%

\subsection{Spatially Separable Attention Mechanism}
\begin{figure}
\centering
\includegraphics[width=2.9in]{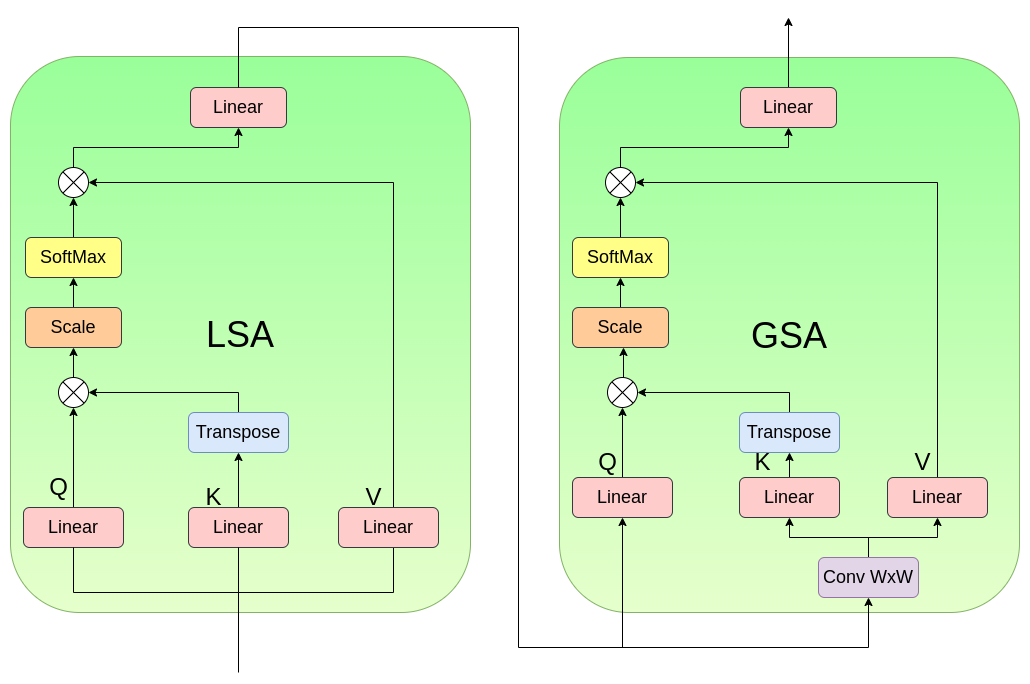}
\captionof{figure}{Spatially Separable Attention Mechanism (Locally grouped self-attention (LSA) followed by Global sub-sampled attention (GSA)). }
\label{attn}
\end{figure}

First, let's summarize the self-attention mechanism of a transformer. Every channel matrix that enters the attention block is fed to three independent linear layers as shown in Fig.{~\ref{attn}}. The outputs of these three branches are queries (Q), keys (K), and values (V) respectively. If the input is $X$, these values are calculated as follows
\begin{figure}
\centering
\includegraphics[width=2.8in, height=1.8in]{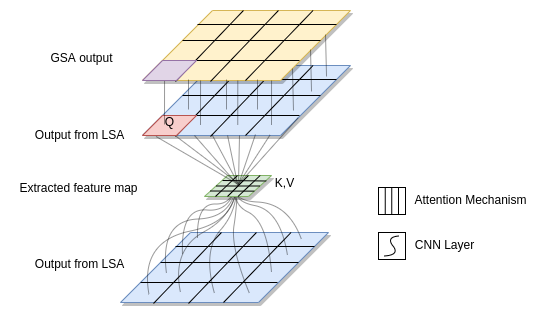}
\captionof{figure}{Global Sub-Sampled Attention (GSA) with sampling performed by a convolutional layer (shown in curved lines) followed by LSA (shown in vertical lines). }
\label{attn1}
\end{figure}
\[Q_{n} = XW^{Q}_{n}, \quad K_{n} = XW^{K}_{n}, \quad V_{n} = XW^{V}_{n}, \]
where $W^{Q}_{n}, W^{K}_{n}, W^{V}_{n}$ are the weights of the respective linear layers on the $n^{th}$ head of a $P$ headed multi-head attention block. With this, the attention is calculated as,
\begin{equation}
    A_{n} = Softmax\bigg(\frac{Q_{n}K_{n}^{T}}{\sqrt{d}}\bigg),
\end{equation}
where $A_{n}$ is the attention on the $n^{th}$ head and $d$ is the output dimension of $Q_{n}$ and $K_{n}$. This attention is then multiplied with the values across all the heads
\begin{equation}
    Y_{n} = A_{n}V_{n},
\end{equation}
which are then concatenated to get the final output.
\begin{equation}
    Y = [Y_{1}, Y_{2}, \cdots, Y_{T}].
\end{equation}
This is also called full attention or global attention as the receptor region of the attention block is the full channel matrix. This global attention mechanism has a complexity of $\mathbb{O}(L^4d)$ when operating on a channel of dimension $L\times L$ with encoded dimension $d$ \cite{ssa}. One way to reduce the complexity is to reduce the receptor region of the attention by using windowed attention, where each channel matrix is sub-divided into $m\times m$ smaller matrices with dimensions $W \times W$, where $W = \frac{L}{m}$ and attention is calculated for each window separately. This is called Locally Grouped Self-Attention (LSA) and this reduces the complexity to $\mathbb{O}(\frac{L^4}{m^4}d)$.

As the windows are fixed and do not communicate with one another, the antenna correlation across windows is now lost and can't be utilized in compressing the channel matrix. To solve this, we can introduce a global attention layer after LSA, but that would only increase the complexity further. So, we introduce another layer of locally grouped attention after LSA that can capture the antenna correlations between windows. This can be achieved by a Global Sub-Sampled Attention (GSA) layer which is shown in Fig.{~\ref{attn1}}. 

In GSA, LSA's output is first followed by a CNN layer (with $stride$ $= W$). The output of this layer is a $m\times m$ feature map in which each element represents a window from which it is extracted. This now becomes keys and values ($K$ and $V$) for another layer of windowed attention whose queries ($Q$) are the same output of LSA that we used to obtain the feature map from as shown in Fig.{~\ref{attn1}}. Suppose $X$ is the output from LSA, we apply a CNN layer to $X$ to get a feature map of dimensions $m\times m$. This feature map becomes $K$ and $V$ for $X$, which becomes the query, $Q$. We are constructing the global attention from the feature map which is a sub-sampled version of $X$, hence the name global sub-sampled attention. This GSA layer has a complexity of $\mathbb{O}(m^2L^2d)$ \cite{ssa} and with this, the total complexity of the attention mechanism becomes $\mathbb{O}(\frac{L^4}{m^4}d)+\mathbb{O}(m^2L^2d)$. The entire attention mechanism (LSA+GSA) is shown in Fig.{~\ref{attn}}. Note that this approach of breaking down a complex operation into two simpler operations is similar to separable convolutions (point-wise + depth-wise) \cite{sepcnn}. Hence, the name spatially separable attention. \par
\subsection{Spatially Separable Attention Transformer Block (STB)}
\begin{figure}
\centering
\includegraphics[width=2.4in,height=2.3in]{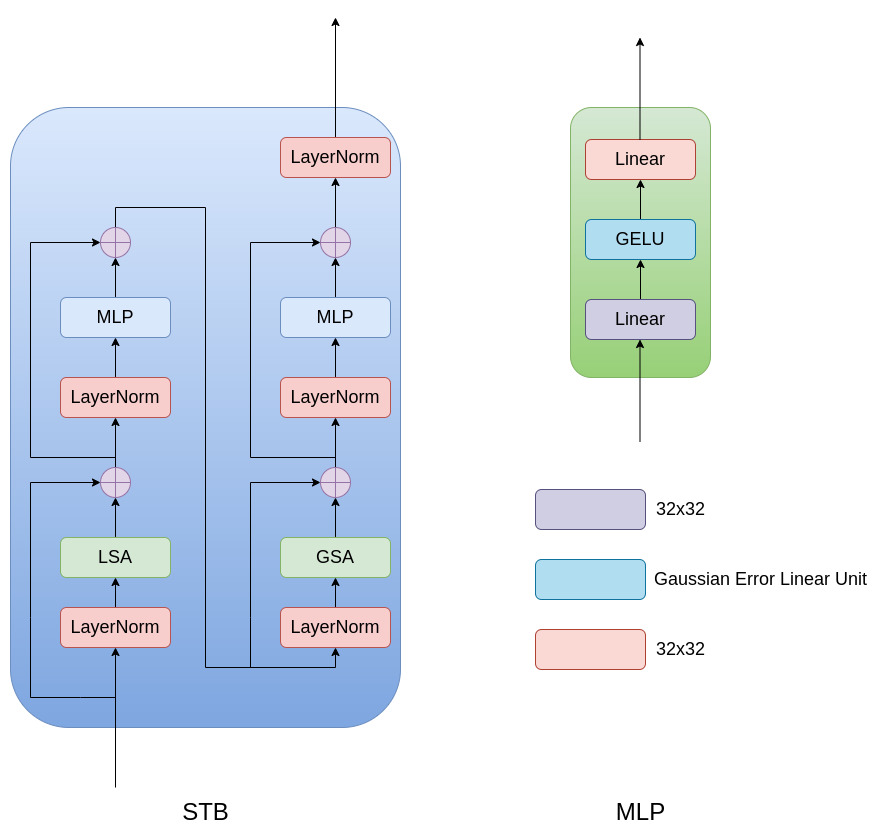}
\captionof{figure}{STB consists of both LSA and GSA each followed by an add $\&$ normalize layer and a Multi-Layer perceptron (MLP). MLP has a linear layer of dimension $32\times 32$ followed by a Gaussian Error Linear Unit (GELU).}
\label{stb}
\end{figure}

STB consists of four different types of blocks which are LSA, GSA, LayerNorm, and Multi-Layer Perceptron (MLP) as shown in Fig.{~\ref{stb}}. %LSA is the same as a windowed multi-head self-attention layer. LSA calculates windowed self attention. On the other hand, GSA constructs global attention from sub-sampled feature maps as described in the last section. 
MLP block has a linear layer followed by a Gaussian Error Linear Unit (GELU) non-linearity and a linear layer again as shown in Fig.{~\ref{stb}}. The entire architecture of STNet with STBs is shown in Fig.{~\ref{net}}. It can be seen that its encoder is slightly more complex than that of CLNet \cite{clnet}, CRNet \cite{crnet}, or CSInet \cite{csinet} with more CNN layers and an STB block. The reason for which is explained in Section IV.

% \subsection{Overall Description}
% The input channel matrix to the encoder is of size $N_{c} \times N_{t} \times 2$. For simplicity, consider $N_{c} = N_{t} = N$. Then, choose a $W$ that perfectly divides $N$. This is the window size of our attention mechanism. There will be $\frac{N^2}{W^2}$ such windows. The output from CNN layers is fed to an STB that performs the spatially separable attention. This output is then given to a couple of convolutional, deconvolutional, and layer normalization layers along with a skip connection. Whose output is reshaped to $2N^2\times 1$ and fed to a fully connected network to compress it to $M\times 1$ which is then transmitted over the air to the gNB. The received compressed vector at gNB is of size $M \times 1$ and is first passed through a linear layer to decompress it to a size of $2N^2\times 1$ which is then reshaped to $N\times N \times 2$. This reshaped channel matrix is now passed to both the CNN stem and the Transformer stem. At the end of the decoder, we use a sigmoid layer to finally reconstruct the channel matrix. 

\begin{figure*}
\includegraphics[width=\linewidth]{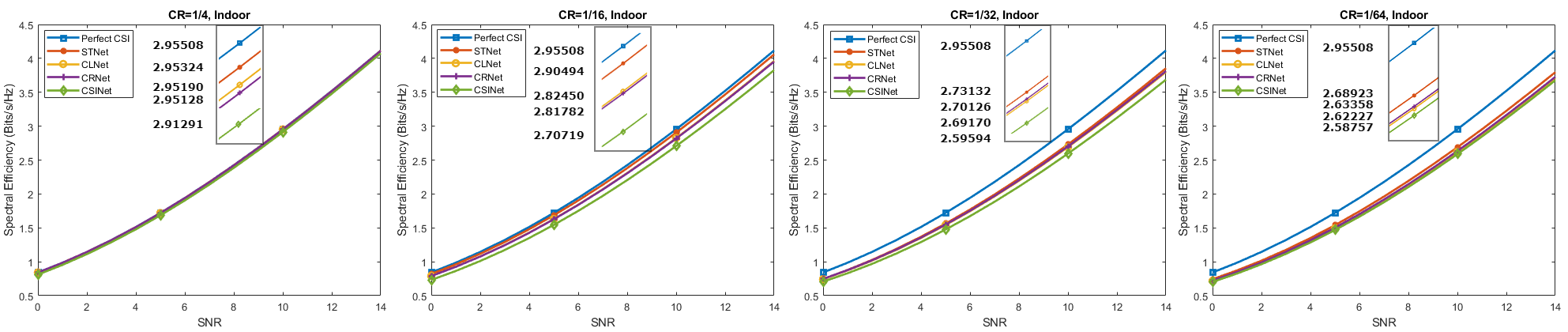}
\captionof{figure}{Spectral Efficiency vs SNR plots for different CSI feedback methods along with perfect CSI. Spectral efficiency values at SNR=10dB are zoomed in and labeled for clarity. }
\label{se}
\end{figure*}

% \begin{figure*}
%      \centering
%      \begin{subfigure}[b]{0.2\textwidth}
%          \centering
%          \includegraphics[width=\textwidth]{512I.png}
%          \caption{}
%          \label{fig:y equals x}
%      \end{subfigure}
%      \hfill
%      \begin{subfigure}[b]{0.2\textwidth}
%          \centering
%          \includegraphics[width=\textwidth]{256I.png}
%          \caption{}
%          \label{fig:three sin x}
%      \end{subfigure}
%      \hfill
%      \begin{subfigure}[b]{0.2\textwidth}
%          \centering
%          \includegraphics[width=\textwidth]{128I.png}
%          \caption{}
%          \label{fig:five over x}
%      \end{subfigure}
%      \hfill
%      \begin{subfigure}[b]{0.2\textwidth}
%          \centering
%          \includegraphics[width=\textwidth]{64I.png}
%          \caption{}
%          \label{fig:five over x}
%      \end{subfigure}
%      \hfill
%      \begin{subfigure}[b]{0.2\textwidth}
%          \centering
%          \includegraphics[width=\textwidth]{32I.png}
%          \caption{}
%          \label{fig:five over x}
%      \end{subfigure}
%         \caption{Three simple graphs}
%         \label{fig:three graphs}
% \end{figure*}

\section{Analysis}
\begin{table*}[h!]
\caption{Performance over Cost 2100 dataset}
\centering
\begin{tabular}{|c|c c|c c|c c|c c|c c|} 
 \hline
  \multicolumn{11}{|c|}{NMSE}\\
 \hline
 Compression Ratio ($\gamma$) & \multicolumn{2}{c}{1/4} & \multicolumn{2}{c}{1/8} & \multicolumn{2}{c}{1/16} & \multicolumn{2}{c}{1/32} & \multicolumn{2}{c|}{1/64}\\
 \hline
 Scenario & Indoor & Outdoor & Indoor & Outdoor & Indoor & Outdoor & Indoor & Outdoor & Indoor & Outdoor \\
 \hline
 CSINet & -17.36 & -8.75 & / & / & -8.65 & -4.51 & -6.24 & -2.81 & -5.84 & -1.93 \\ 
 \hline
 CRNet & -26.99 & -12.71 & -16.01 & -8.04 & -11.35 & -5.44 & -8.93 & -3.51 & -6.49 & -2.22 \\
 \hline
 CLNet & -29.16 & -12.88 & -15.60 & -8.29 & -11.15 & -5.56 & -8.95 & -3.49 & -6.34 & -2.19  \\
 \hline
 CSIFormer & / & / & / & / & / & / & -9.32 & -3.51 & -6.85 & -2.25\\
 \hline
 TransNet & \textbf{-32.38} & \textbf{-14.86} & \textbf{-22.91} & \textbf{-9.99} & -15.00 & \textbf{-7.82} & \textbf{-10.49} & \textbf{-4.13} & -6.08 & \textbf{-2.62} \\
 \hline
 STNet & -31.81* & -12.91* & -21.28* & -8.53* & \textbf{-15.43} & -5.72* & -9.42* & -3.51* & \textbf{-7.81} & -2.46*  \\
 \hline \hline
 \multicolumn{11}{|c|}{FLOPS and RUNTIME (in seconds)}\\
 \hline
  Compression Ratio ($\gamma$) & \multicolumn{2}{c}{1/4} & \multicolumn{2}{c}{1/8} & \multicolumn{2}{c}{1/16} & \multicolumn{2}{c}{1/32} & \multicolumn{2}{c|}{1/64}\\
 \hline
  Scenario & FLOPS & Runtime & FLOPS & Runtime & FLOPS & Runtime & FLOPS & Runtime & FLOPS & Runtime \\ %[0.5ex] 
   \hline
%  Scenario & Indoor & Outdoor & Indoor & Outdoor & Indoor & Outdoor & Indoor & Outdoor & Indoor & Outdoor\\
%  \hline
  CSINet & 5.41M & 0.0001 & 4.37M & 0.0001 & 3.84M & 0.0001 & 3.58M & 0.0001 & 3.45M & 0.0001 \\ 
 \hline
 CRNet & 5.12M & 0.0003 & 4.07M & 0.0003 & 3.55M & 0.0003 & 3.28M & 0.0003 & 3.16M & 0.0003 \\
 \hline
 CLNet & 4.42M & 0.0002 & 3.37M & 0.0002 & 2.85M & 0.0002 & 2.58M & 0.0002 & 2.45M & 0.0002 \\
 \hline
 CSIFormer & / & - & / & - & / & - & 5.41M & - & 5.54M & - \\
 \hline
 TransNet & 35.72M & - & 34.70M& - & 34.14M & - & 33.88M & - & 33.75M & - \\
 \hline
  STNet & 5.22M & 0.0004 & 4.38M & 0.0003 & 3.96M & 0.0003 & 3.75M & 0.0003 & 3.65M & 0.0003\\
 \hline
   CSITransformer & / & / & / & / & / & 0.003 & / & 0.002 & / & /\\
    \hline
\end{tabular}\\
/ indicates that the performance is not reported in the original paper\\
* indicates the second-best value in that column and 
- indicates that the code is not made public in order to generate the results
\end{table*}
\vspace{-0.15cm}
% \begin{table*}[h!]
% \caption{Performance over UMa dataset}
% \centering
% \begin{tabular}{|c|c|c|c|c|c|} 
%  \hline
%   \multicolumn{6}{|c|}{NMSE}\\
%  \hline
%  Compression Ratio ($\gamma$) & 1/4 & 1/8 & 1/16 & 1/32 & 1/64\\
%  \hline
%  CSINet & / & / & / & / & /\\ 
%  \hline
%  CRNet & / & / & / & / & /\\
%  \hline
%  CLNet & -15.28 & -15.10 & / & / & / \\
%  \hline
%  STNet & -16.01 & -15.12 & -10.68 & -7.99 & -5.63  \\
%  \hline
% \end{tabular}\\
% / - indicates that the performance is not reported in the original paper\\
% * - indicates it is the second best value in that column
% \end{table*}
\subsection{Model Performance}
We consider a system with $32\times1$ antennas (i.e., 32 antennas at BS and 1 antenna at UE). For evaluation purposes, we choose the COST2100 dataset with two scenarios: the indoor picocellular scenario at 5.3GHz and the outdoor rural scenario at 300MHz. We choose $N_{c} = 32$, window size: $W=8$ and number of heads of multi-head attention: $P=4$. The training, validation, and test datasets consist of  100,000, 30,000, and 20,000 matrices, respectively. Batch size is set to 200 and epochs to 1000. The learning rate is 0.001 and the loss function is the Mean Squared Error (MSE) with an Adam optimizer.
\begin{equation}
    MSE = \frac{1}{B}\sum_{i=1}^{B} ||\bm{H}-\bm{\hat{H}}||^2,
\end{equation}
where $\bm{H}$ is the input channel matrix, $\bm{\hat{H}}$ is the reconstructed channel matrix and $B$ is the batch size. We use Normalised Mean Square Error (NMSE) as the performance metric which is defined as follows
\begin{equation}
    NMSE = \mathbb{E} \bigg\{\frac{||\bm{H}-\bm{\hat{H}}||^2}{||\bm{H}||^2}\bigg\}.
\end{equation}
%\begin{equation}
   % \rho = \mathbb{E} \bigg\{\frac{1}{N_{c}}\sum_{n=1}^{N_{c}}\frac{|\hat{h}^{H}h_{n}|}{||\hat{h}_{n}||||h_{n}||}\bigg\}
%\end{equation}
% where $N_{c}$ is the number of sub-carriers in the truncated matrix, $h_{n}$ and $\hat{h_{n}}$ are the the channel vectors and reconstructed channel vectors at BS on the $n^{th}$ sub-carrier respectively. Cosine similarity captures the efficiency of a beamforming vector. If $v_{n} = \frac{\hat{h_{n}}}{||\hat{h_{n}}||}$ then the received channel at UE is $\frac{\hat{h}^{H}h_{n}}{||\hat{h_{n}}||}$. 
The number of floating-point operations per second (FLOPS) and runtime delay of the model is other important factors when comparing the models as deployment is also done in UEs which are power and memory-constrained devices. So we tabulated the NMSE results of our model over the COST2100 dataset and its FLOPs and runtimes compared with various other models in Table I. \par
STNet is compared with the recently proposed transformer-based models CSIFormer \cite{csifromer} and TransNet \cite{transnet}. TransNet may have performed well in most cases but it takes significantly more FLOPs to achieve that. For the indoor case with $CR$ $=1/16$, STNet achieved $102.86\%$ of the performance of TransNet with just $11.6\%$ of FLOPs. Similarly, for $CR$ $=1/64$, STNet achieved $128.45\%$ of the performance of TransNet with just $10.8\%$ of its FLOPs. Also, notice that STNet performs better than CSIFormer in every case while consuming less number of FLOPs. For instance consider outdoor environment with $CR$ $=1/64$. STNet achieved $116.44\%$ of the performance of CSIFormer with $65.88\%$ of its FLOPs. Also from Table I, we can see that the runtimes of STNet are comparable to other models. It is evident from these results that STNet exploits the trade-off between performance and complexity perfectly. \par

% The number of floating-point operations per second (FLOPS) and runtime delay of the model are another important factors when comparing the models as deployment is also done in UEs which are power and memory-constrained devices. So we tabulated the NMSE results of our model over the COST2100 dataset and its FLOPs and delays compared with various other models in Table I. \par
% STNet is compared with the recently proposed transformer-based models CSIFormer \cite{csifromer} and TransNet \cite{transnet}. TransNet may have performed well in most cases but it takes significantly more FLOPs to achieve that. For an indoor scenario with $CR$ $=1/4$, STNet achieved $97.24\%$ of the performance of TransNet with just $14.6\%$ of its FLOPs. For the indoor case with $CR$ $=1/16$, STNet even achieved $102.86\%$ of the performance of TransNet with just $11.6\%$ of FLOPs. Similarly, for the indoor case with $CR$ $=1/64$, STNet achieved $128.45\%$ of the performance of TransNet with just $10.8\%$ of its FLOPs. Also, notice that STNet performs better than CSIFormer in every case while consuming less number of FLOPs. For instance consider outdoor environment with $CR$ $=1/64$. STNet achieved $116.44\%$ of the performance of CSIFormer with $65.88\%$ of its FLOPs. Also from Table I, we can see that the runtimes of STNet are comparable to other models. It is evident from these results that STNet exploits the trade-off between performance and complexity perfectly. \par 
Although the runtimes of CSIFormer and TransNet are not available, the runtimes of a similar full attention mechanism based model called CSITransformer \cite{empower} are available which can be used for comparison with STNet. CSITransformer is evaluated on a different dataset so its NMSE results are not listed. However, the size of the channel matrices used by it is $32\times32$ which is the same as all the other models so the runtimes of it can be used in a fair comparison with STNet. Thus, the runtimes of CSITransformer for $CR$ $=1/16$ and $CR$ $=1/32$ are listed in Table I and we can see that STNet is faster than CSITransformer in both scenarios.
% \vspace{-0.45cm}
% However, the runtimes of CRNet are higher than CSINet despite having fewer FLOPs because FLOPs doesn't necessarily capture the computational efficiency of the model. Two models with same FLOPs can have different runtimes depending on the number of parallelizable operations each model has. The model with most parallelizable operations will have the least runtime despite having more FLOPs. In the case of CSINet \cite{csinet} and CRNet \cite{crnet}, although CRNet has fewer FLOPs than CSINet, the latter achieved lesser delays. It is because CRNet uses CRBlocks (Fig.{~\ref{crblock}}) which has two stems where the second stem depends on the first one. As the stems are not independent parallelization is not possible thus increasing the runtime. Similarly, STNet also has two stems where the transformer stem waits for the CNN stem to finish as the output of the latter is utilized by the former. This increases the runtime of STNet but it's low complexity means that it can be efficiently implemented on limited hardware.

\subsection{Communication System Performance}
A Massive MIMO system can achieve high capacities by using transmit precoding. Therefore, we use the widely common linear Zero-Forcing (ZF) transmit precoding to evaluate the overall performance improvement of the communication system due to different CSI feedback methods \cite{csifromer}. For evaluation, the spectral efficiency of each method is plotted against SNR for various compression ratios as shown in Fig.{~\ref{se}}. At 10dB SNR, spectral efficiency values of all the methods are labeled and from Fig.{~\ref{se}}, we can conclude that STNet performs better than every model in every scenario and the difference is more profound for $CR$ $=1/16$ as STNet achieves lowest NMSE for this scenario. This improvement in system performance is achieved by STNet's ability to capture antenna and sub-carrier correlations at the encoder side. This operation is so critical that the encoder consumes more than $40\%$ of the entire STNet's resources. For example in an indoor scenario with $CR$ $=1/4$, STNet's encoder has 2.09 Million FLOPs which is $40.03\%$ of the total FLOPs which is 5.22 Million. On the other hand, CLNet's encoder under similar conditions has 1.11 Million FLOPs which is $25.11\%$ of the total FLOPs which is 4.42 Million. Although STNet has a slightly higher encoder complexity than CSINet, CRNet, or CLNet, its encoder complexity would still be better than other transformer-based models making it a promising choice for storage and computational limited applications because of its higher spectral efficiency and lower or similar runtime as other models. 
\section{Conclusion}
In this work, a lightweight transformer architecture with spatially separable attention is introduced for CSI feedback. Along with this, a hybrid  approach to channel reconstruction is also introduced where we use a two stems approach (CNN and Transformer) that improves the channel reconstruction quality. We evaluated the performance and runtime of STNet along with other models on the COST2100 dataset. Combining both techniques, STNet produced the best performance per floating-point operation among various other models.

\bibliography{reference}
\bibliographystyle{ieeetr}
\end{document}